# Confinement Effects of External Fields and Topological Defect on Hydrogen Atom in a Quantum-plasma Environment

*C. O. Edet[a,b] and A. N. Ikot[b]

[a]Department of Physics, Cross River University of Technology, Calabar, Nigeria
[b]Theoretical Physics Group, Department of Physics, University of Port Harcourt, Choba, Nigeria.

*Email for correspondence: collinsokonedet@gmail.com

## ABSTRACT

This study looks at the confinement effects of Aharonov-Bohm (AB) flux and magnetic fields, as well as topological defects in a quantum plasma, on the hydrogen atom. The joint effects show that the system is extremely attractive. Furthermore, as we've shown, the joint effect of the fields is greater than the sum of the individual effects, resulting in a significant change in the system's bound state energy. The magnetic field can be used as a control parameter or booster, whereas the topological defect and AB field are needed to hold the hydrogen atom in quantum plasmas at a low energy. The findings of our research may be extended to atomic structure and plasma collisions.

## 1. INTRODUCTION

Potential models have been used to model several phenomena right from the early days of quantum mechanics and its related disciplines. It has been used to model interatomic interactions in diatomic molecules, quantum dots, quakonium interactions, plasma etc [1]. Debye and Huckel suggested a theory [2] in 1923 that plasmas could be described theoretically with the well-known screened Coulomb potential. This model or potential provides a cutting-edge plasma treatment through the screening effect. This model is provided to simulate the plasma screening effect for weakly coupled plasmas. It is given by [3]; $V(r) = \frac{(Ze^2)e^{(-r/\lambda_D)}}{r}$, where $\lambda_D$ represents the Debye length or Debye screening parameter and determines the interaction between electrons in Debye plasma. This potential is employed to account for pair correlations.

A thoughtful presentation by Shukla and Eliasson [4] showed that the modified Debye-Huckel potential also known as the exponential-cosine-screened Coulomb potential $V(r) = -\left(Ze^2/r\right)\exp\left(-r/\lambda_D\right)\cos\left(r/\lambda_D\right)$ can be used to model the effective screened potential of a test charge of mass $m$ in a dense quantum plasma. It was also established that $V(r) = -(Ze^2/r)\exp(-r/\lambda_D)\cos(gr/\lambda_D)$ can be used to model weakly coupled plasmas with $g = 0$ and dense quantum plasmas with $g = 1$ [3].

In another consideration, Yoon and Yun [5] proposed that the well-known Hulthen potential [6]; $V_H(r) = -\dfrac{Ze^{-r/\lambda_D}}{\lambda_D\left(1 - e^{-r/\lambda_D}\right)}$ can be used for describing the plasma-screening effect on the continuum electron capture rate in the solar core region. The advantage of this potential, among many others, is that it provides us with analytic solutions, allowing us to quickly and accurately analyze reaction rates and a variety of other observables. This model has been used to describe the plasma screening effect on a hydrogen atom embedded in weakly coupled plasma by Bahar *et al.* [7]. Myriads of research in the literature have concentrated on researching the influence of various fields on the hydrogen atom surrounded by plasmas [8] and references therein.

When plasma density is intensified adequately, quantum effects become very fascinating. This includes degeneracy effects, which becomes important when $T \ll T_F$, where $T$ is the plasma temperature, $T_F = \dfrac{E_F}{k_B}$ is the Fermi temperature, defined in Fermi energy as $E_F = (\hbar^2/2m)(3\pi^2 n_0)^{2/3}$, where $k_B$ is Boltzmann's constant and $\hbar$ is the reduced Planck constant. In this model, particle dispersive effects tend to be important for short scale-lengths (comparable to the characteristic de-Broglie length) when $\hbar\omega_p/k_B T_F \approx 1$, where $\omega_p$ is the plasma frequency. Moreover, in quantum kinetic theory, these effects can be properly modeled using the perturbation theory.

In plasmas, in order to further improve these models, the electron spin is taken into account, which introduces a magnetic dipole force, spin precession, and spin magnetization currents into the picture [9]. However, in the present consideration the vector potential comprising of magnetic and Aharanov-Bohm fields and topological defect are introduced.

In fact, there has been an increasing interest in plasmas of low-temperature and high densities, where quantum properties tend to be important [10]. Promising applications include spintronics [11], and plasmonics [12]. Quantum plasma effects can also be of interest in experiments with solid density targets [13]. Important classifications of dense plasmas include whether they are strongly or weakly coupled, and whether they are degenerate or nondegenerate [9].

We have sufficient resources to research the hydrogen atom in weakly coupled quantum plasmas under the influence of an Aharonov-Bohm (AB) flux field and a uniform magnetic field with topological defects, based on the information we have gathered. In non-relativistic quantum mechanics and quantum field theory, the hydrogen atom stands out as a straightforward dual-body problem with an empirical closed form solution [3, 14,15]. When studying quantum effects in very complicated structures, a thorough understanding of its basic structure is critical.

Numerous papers [7-18, 16] have investigated the effects of an external electric and magnetic field on the hydrogen atom. We propose that, in addition to using AB and magnetic fields to regulate the energy levels or localization of a hydrogen atom's quantum state in quantum plasmas, the topological defect could be used.

We show in this paper that the effect topological defect and other external fields have an effect. This paper is an extension of the works of Bahar *et al.* [7] and Falaye *et al.* [8]. Both articles considered the Hydrogen atom in Minkwoski space but in the present study, we have taken a step forward to consider the effect of a defect in the medium. In various areas of physics, such as gravitation [17], condensed matter physics [18], etc, the study of the effect of topology on the physical properties of various systems has been of critical relevance and this has been greatly studied by several authors [19-37]. Hydrogen atom has been studied in a medium with topological defect by several authors [38-39].

As a result, it's critical to investigate the effect of disclination on hydrogen atoms in a quantum plasma system with AB and magnetic fields. However, we would like to conduct this research using the Hulthen potential model of the form;

$$V_H(r) = -\frac{Ze^{-r/\lambda_D}}{\eta\lambda_D\left(1-e^{-r/\lambda_D}\right)} \tag{1}$$

where the parameter $\eta$ is a consequent of the topological defect. The effects of these two external fields on a hydrogen atom with a topological defect in a weakly coupled quantum plasma was investigated in this article. As a result, we expect this research would be useful in fields such as atomic structure and plasma collisions, as well as astrophysics.

2. **THEORY AND CALCULATIONS**

Consider the hydrogen atom in a quantum-plasma system, where the AB and $\vec{B}$ fields are all-pervasive. Let's presume there's a disclination or topological defect in this region. The disclination is defined by the line unit [27, 39, 40].

$$ds^2 = dr^2 + \eta^2 r^2 d\phi^2 + dz^2 \tag{2}$$

where $0 < \eta < 1$ is the parameter associated with the deficit of angle. The parameter $\eta$ is related to the linear mass density $\tilde{\mu}$ of the string via $\eta = 1 - 4\tilde{\mu}$. We see here that the azimuthal angle is denoted in the interval $0 \leq \phi \leq 2\pi$ [39].

The Schrodinger equation for this consideration is given as follows, bearing in mind the presence of the perturbing fields and defect:

$$\left(i\hbar\vec{\nabla} - \frac{e}{c}\vec{A}_\eta\right)^2 \psi(r,\phi,z) = 2\mu\left[E_{nm} + \frac{Ze^{-r/\lambda_D}}{\eta\lambda_D\left(1-e^{-r/\lambda_D}\right)}\right]\psi(r,\phi,z), \tag{3}$$

where $E_{nm}$ represents the energy level, $\mu$ is the effective mass of the system. The Laplacian used is defined as follows with Topological defect as [39, 40]. The vector potential, denoted by "$\vec{A}$", is the structure of the vector potential that produces a uniform magnetic field in conical space. It is given as: $\vec{A} = \left(0, \dfrac{\vec{B} e^{-r/\lambda_D}}{\eta\left(1 - e^{-r/\lambda_D}\right)} + \dfrac{\phi_{AB}}{2\pi r}, 0\right)$. The parameter $\eta$ represents the influence of the topological defect in the region. We assume a wave function in cylindrical coordinates as $\psi(r,\phi) = \dfrac{1}{\sqrt{2\pi r}} e^{im\phi} H_{nm}(r)$, where $m$ denotes the magnetic quantum number. On inserting this wave function and the vector potential into Eq. (3), we get:

$$H''_{nm}(r) + \dfrac{2\mu}{\hbar^2}\left[E_{nm} - V_{eff}(r)\right] H_{nm}(r) = 0 \tag{4}$$

where $V_{eff}(r)$ is the effective potential defined as follows;

$$V_{eff}(r) = \dfrac{Ze^{-r/\lambda_D}}{\eta \lambda_D \left(1 - e^{-r/\lambda_D}\right)} + \hbar\omega_c \left(\dfrac{m}{\eta^2} + \dfrac{\xi}{\eta}\right) \dfrac{e^{-r/\lambda_D}}{\left(1 - e^{-r/\lambda_D}\right) r} + \left(\dfrac{\mu\omega_c^2}{2}\right) \dfrac{e^{-2r/\lambda_D}}{\left(1 - e^{-r/\lambda_D}\right)^2} + \dfrac{\hbar^2}{2\mu}\left[\dfrac{\left(\dfrac{m}{\eta^2} + \xi\right)^2 - \dfrac{1}{4}}{r^2}\right] \tag{5}$$

where $\xi = \dfrac{\phi_{AB}}{\phi_0}$ is an integer with the flux quantum $\phi_0 = \dfrac{hc}{e}$ and $\omega_c = \dfrac{e\vec{B}}{\mu c}$ denotes the cyclotron frequency. Employing the approximation proposed by Greene and Aldrich [42] defined as follows; $\dfrac{1}{r^2} = \dfrac{1}{\lambda_D^2 \left(1 - e^{-r/\lambda_D}\right)^2}$. This approximation have been widely used by several authors [43-45] to overcome such kind of barriers. With the following substitution $\rho = e^{-r/\lambda_D}$ into Eq. (4), we can simply write Eq. (4) in the $\rho$-coordinate as follows;

$$\dfrac{d^2 H_{nm}(\rho)}{d\rho^2} + \dfrac{1}{\rho} \dfrac{d H_{nm}(\rho)}{d\rho} + \dfrac{1}{\rho^2 (1-\rho)^2}\left[\begin{array}{l}-(\varepsilon_{nm} + \delta + \beta_2)\rho^2 + (2\varepsilon_{nm} - \beta_1 + \delta)\rho \\ -(\varepsilon_{nm} + m')\end{array}\right] H_{nm}(\rho) = 0 \tag{6}$$

where we have introduced the following dimensionless notations;

$$-\varepsilon_{nm} = \dfrac{2\mu \lambda_D^2 E_{nm}}{\hbar^2},\ \delta = \dfrac{2\mu \lambda_D Z}{\eta \hbar^2},\ \beta_1 = \dfrac{2\mu \lambda_D \omega_c}{\hbar}\left(\dfrac{m}{\eta^2} + \dfrac{\xi}{\eta}\right),\ \beta_2 = \dfrac{\mu^2 \lambda_D^2 \omega_c^2}{\hbar^2},\ m' = \left(\dfrac{m}{\eta} + \xi\right)^2 - \dfrac{1}{4} \tag{7}$$

for mathematical simplicity. We assume the following ansatz of the form;

$$H_{nm}(\rho) = \rho^\sigma (1-\rho)^\Lambda f_{nm}(\rho) \tag{8}$$

where

$$\sigma = \sqrt{\varepsilon_{nm} + \gamma}$$
$$\Lambda = \frac{1}{2} + \sqrt{\frac{1}{4} + \beta_2 + \beta_1 + m'} \qquad (9)$$

On substitution of Eq. (8) into Eq. (7), we obtain a hypergeometric differential equation and by extension the quantization condition as follows:

$$\sigma_n = -n - \Lambda \pm \sqrt{\varepsilon_{nm} + \delta + \beta_2} \qquad (10)$$

On making $\varepsilon_{nm}$ the subject of the expression by carrying some simple manipulative algebra, we get;

$$\varepsilon_{nm} = -m' + \frac{1}{4}\left[\frac{\left(n + \frac{1}{2} + \sqrt{\frac{1}{4} + \beta_2 + \beta_1 + m'}\right)^2 - \delta - \beta_2 + m'}{\left(n + \frac{1}{2} + \sqrt{\frac{1}{4} + \beta_2 + \beta_1 + m'}\right)}\right]^2 \qquad (11)$$

As a result, if the value of the dimensionless parameters in Eq. (7) is substituted into Eq. (11), the following solutions are obtained:

$$E_{nm} = \frac{\hbar^2 m'}{2\mu\lambda_D^2} - \frac{\hbar^2}{8\mu\lambda_D^2}\left[\frac{\left(n + \frac{1}{2} + \sqrt{\frac{1}{4} + \frac{\mu^2\omega_c^2\lambda_D^2}{\hbar^2} + \frac{2\mu\omega_c\lambda_D}{\hbar}\left(\frac{m}{\eta^2} + \frac{\xi}{\eta}\right) + m'}\right)^2 - \frac{2\mu\lambda_D Z}{\eta\hbar^2} - \frac{\mu^2\omega_c^2}{\hbar^2\eta^2} + m'}{\left(n + \frac{1}{2} + \sqrt{\frac{1}{4} + \frac{\mu^2\omega_c^2\lambda_D^2}{\hbar^2} + \frac{2\mu\omega_c\lambda_D}{\hbar}\left(\frac{m}{\eta^2} + \frac{\xi}{\eta}\right) + m'}\right)}\right]^2 \qquad (12)$$

For the sake of completeness, we continue to find the system's wave function [46,47]. The corresponding unnormalized wave function is obtain as

$$H_{nm}(\rho) = C_2 \frac{\Gamma(2\sigma + 1 + n)}{\Gamma(2\sigma + 1)} \rho^\sigma (1-\rho)^\Lambda (-1)^n {}_2F_1\left(-n, 2(\sigma + \Lambda) + n; 2\sigma + 1; \rho\right) \qquad (13)$$

The corresponding wave function is obtain as

$$\psi(r,\phi) = \frac{1}{\sqrt{2\pi r}} C_2 \frac{\Gamma(2\sigma + 1 + n)}{\Gamma(2\sigma + 1)} \times$$
$$(-1)^n e^{im\phi} \left(e^{-r/\lambda_D}\right)^{\sqrt{\varepsilon_{nm} + \gamma}} \left(1 - e^{-r/\lambda_D}\right)^{\frac{1}{2} + \sqrt{\frac{1}{4} + \beta_2 - \beta_1 + \beta_3 + \gamma}} {}_2F_1\left(-n, n + 2(\sigma + \Lambda); 2\sigma + 1; e^{-r/\lambda_D}\right) \qquad (14)$$

## 3. DISCUSSIONS

Table 1 shows the eigenvalues of the Hydrogen atom in the presence and absence of external fields (the magnetic field and the AB flux field) as well as topological defects in a.u. and low vibrational $n$ and rotational $m$. From the table, we see that when the external fields and defect are absent (i.e., when $\vec{B} = \xi = 0$ and $\alpha = 1$), the space between the energy levels of the effective potential is narrow and upsurges with rising $n$. It is observed that the degeneracy is found among some states, but on applying the magnetic field, its strength does not only raise the energy levels of the effective potential and spacings between states but transforms the degeneracies to quasi-degeneracies as well. Furthermore, the quasi-degeneracies amongst the states are also eliminated and the energy values is lifted up.

The energy values are reduced and degeneracies are removed when the system is only exposed to the AB flux field, but the pseudo-degeneracies among the states are unaffected. As the quantum number $n$ increases for fixed $m$, the energy levels become more positive and the system becomes strongly positive. When only the topological defect is present, the degeneracies and pseudo-degeneracies are unaffected, and the total interaction potential becomes more attractive. The energy level is also shifted upward by the topological defect.

In the absence of a topological defect, the combined effect of the magnetic and AB fields is greater than the individual effects and the absence of the fields. There is a significant change in energy levels, which we notice. When there is only a magnetic and topological defect, the effect is also great, and there is a significant shift in energy levels, but it is less than when the fields are present alone. When there is only the AB field and a topological defect, the effect is also great, with a significant shift in energy levels, but it is less pronounced than in the previous two cases. The total impacts suggest that the system it is highly repulsive. Furthermore, the mutual effect (triad) of the fields and topological defect is greater than the individual and dual effects, resulting in a significant shift in the system's bound state energy.

Fig. 1 shows the plot of the effective potential energy to simulate weakly coupled quantum plasmas environment with rotational $(m=1)$ levels for various values of $\lambda_D$ with $\xi=6$, $\eta=0.004$, and $\vec{B}=6$. We see that increasing the Debye length and other parameters constant leads to a corresponding increment in the effective potential function. Thus, the potential energy becomes more repulsive. In fig. 1 (b), various values of $\eta$ with $\xi=6$, $\lambda_D=10$, and $\vec{B}=6$. When the intensity of the defect is increased while the other fields remain unchanged, the effective potential function decreases. As a result, potential energy becomes more repulsive. In Fig. 1(c), various values of $\vec{B}$ with $\lambda_D=10$, $\eta=0.004$ and $\xi=6$. When the magnetic field strength is increased while the other fields remain unchanged, the effective potential feature increases in proportion. As a result, potential energy becomes more repulsive. In fig. 1(d), various values of $\xi$ with $\vec{B}=6$, $\xi=6$, and $\eta=0.004$. Increasing the strength of the AB field increases attractiveness of the effective potential.

Fig. 2 shows the effective potential energy to simulate weakly coupled quantum plasmas environment with rotational $(m=1)$ levels. Fig.2 (a), various values of $\lambda_D$ with $\eta=0.004$. The effective potential increases as the inter-nuclear distance increases. In fig. 2(b), same as (a) but with $\eta=1$. The behaviour remains the same, indicating that the presence or removal of the defect has little impact on the result. In fig. 2(c), the approximation used is plotted to show the validity. The approximation is only valid for large $\lambda_D$. For this reason, we choose $\lambda_D \geq 20$ where necessary in our computation. We see that for large $\lambda_D$, the centrifugal term is well-approximated.

Fig. 3 shows the variation of energy values for the hydrogen atom in quantum plasmas and under the influence of the magnetic field and the AB flux field with topological defect in atomic units using the fitting parameters $m = n = 0$ and $\lambda_D = 20$. In fig. 3(a), when the energy is plotted as a function of external magnetic field in the region $0 < \vec{B} < 5$ with various $\xi$ and a fixed $\eta = 0.04$. The energy spectra first rises to a maximum and then reduces again. In addition, the energy increases as the AB field increases. Again in fig. 3(b), when $m = -1$ and $n = 2$, we see that the energy increases linearly as the magnetic field increases. In fig. 3 (c), when $m = n = 0$ and with $\eta = 1$. The energy spectra first rises to a maximum and then reduces again. In addition, the energy increases as the AB field increases.

Fig. 4 shows the variation of energy values for the hydrogen atom in quantum plasmas and under the influence of the magnetic field and the AB flux field with topological defect in atomic units using the fitting parameters $m = n = 0$ and $\lambda_D = 20$. In fig. 4**(a)**, the energy is plotted as a function of the topological defect with various $\xi$ and a fixed magnetic field, $\vec{B} = 1$. We see that the spectrum exhibits saturation. This means that the energy upsurges monotonically as the topological defect approaches unity $(\eta \to 1)$. In fig. 4(b), the energy is plotted as a function of external magnetic field with various $\eta$ and fixed AB field, $\xi = 2$. The energy rises as the magnetic field rises but slightly drops immediately after it reaches a maximum. As shown in Fig. 4 (c), we see that as the magnetic field increases, the energy decreases. The effect of the AB field is great when small. For instance, when $\xi = 4$ is greater than when $\xi = 6$ respectively, we notice that the energy level declines as the magnetic field increases in the interval $0 < B < 5$.

Fig. 5 shows the magnetic field versus inter-particle distance. The magnetic field is seen to increase with increasing inter-particle distance monotonically.

## 4. CONCLUSION

For the first time in a quantum plasmas environment, we have considered the effects of the AB flux, uniform magnetic field, and topological defect on the hydrogen atom in this research paper. The complete effects indicate that the system is highly attractive, while quantum level localizations alter and eigenvalues decrease. Furthermore, as we've shown, the total effect of the fields is greater than the sum of the individual effects, resulting in a large change in the system's bound state energy. Although the magnetic field will act as a controller or boost, the topological defect and AB field are needed to keep the hydrogen atom in quantum plasmas at a low energy. The findings of our research may be extended to atomic structure and plasma collisions.

**Funding**

This research received no external funding.

**Acknowledgements**

C. O. Edet dedicates this work to his Late Father (Mr. Okon Edet Udo). In addition, C. O. Edet acknowledges eJDS (ICTP).

**Data Availability**

The datasets used and/or analyzed during the current study are available from the corresponding author on reasonable request.

**Conflicts of Interest**

The authors declare no conflict of interest.

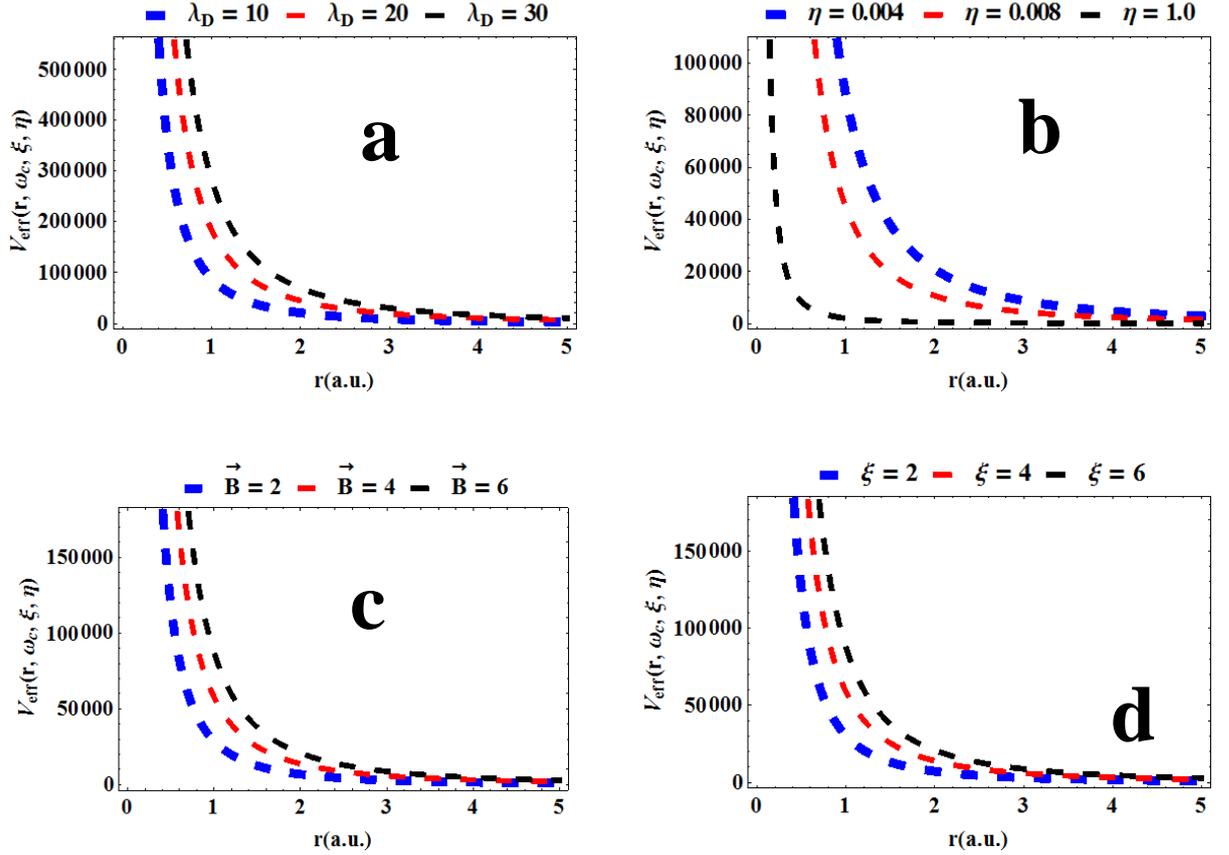

**FIG. 1**. Effective potential energy to simulate weakly coupled quantum plasmas environment with rotational $(m=1)$ levels for **(a)** various values of $\lambda_D$ with $\xi=6$, $\eta=0.004$, and $\vec{B}=6$. **(b)** Various values of $\eta$ with $\xi=6$, $\lambda_D=10$, and $\vec{B}=6$. **(c)** Various values of $\vec{B}$ with $\lambda_D=10$, $\eta=0.004$ and $\xi=6$. **(d)** Various values of $\xi$ with $\vec{B}=6$, $\xi=6$, and $\eta=0.004$. All our computations are in a.u.

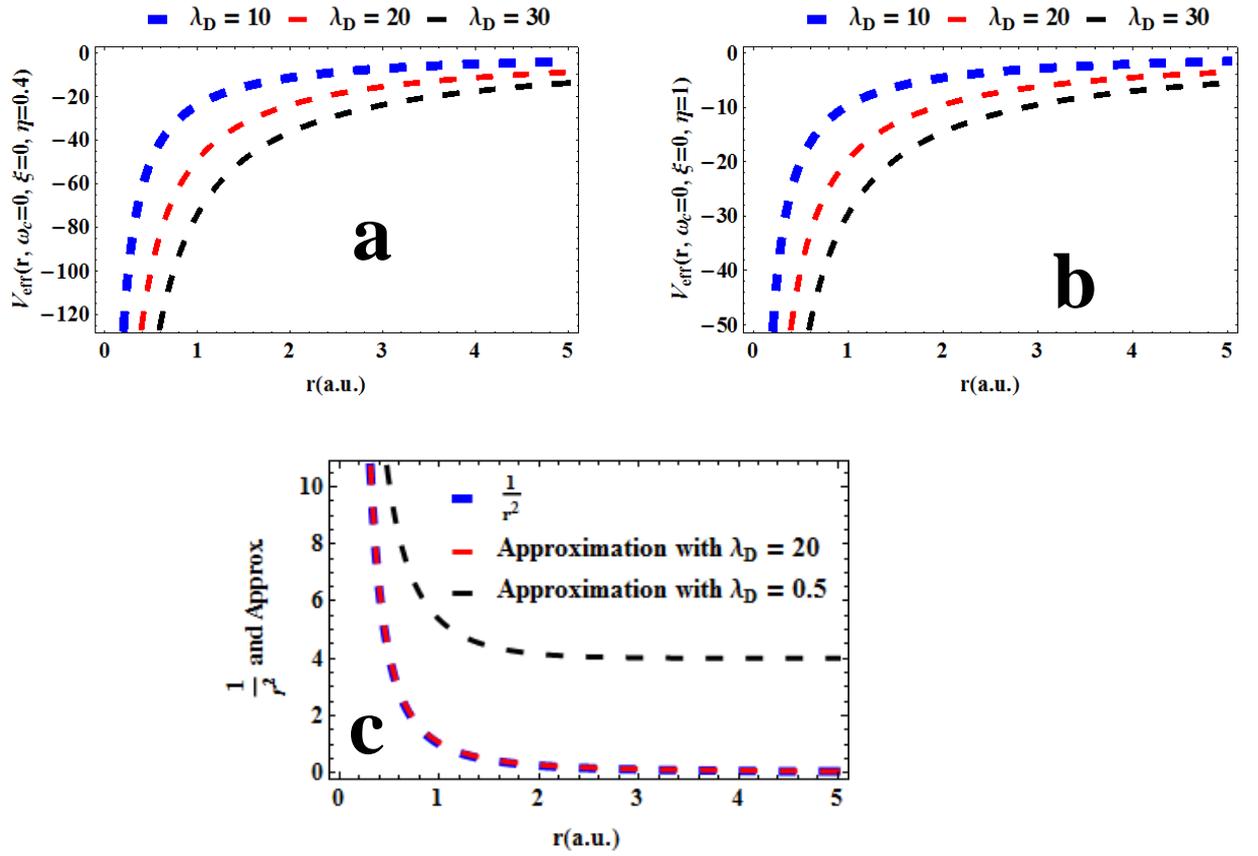

**FIG. 2**. Effective potential energy to simulate weakly coupled quantum plasmas environment with rotational $(m=1)$ levels for **(a)** for various values of $\lambda_D$ with $\eta=0.004$. **(b)** same as **(a)** but with $\eta=1$. **(c)** The approximation used is plotted to show the validity. All our computations are in a.u.

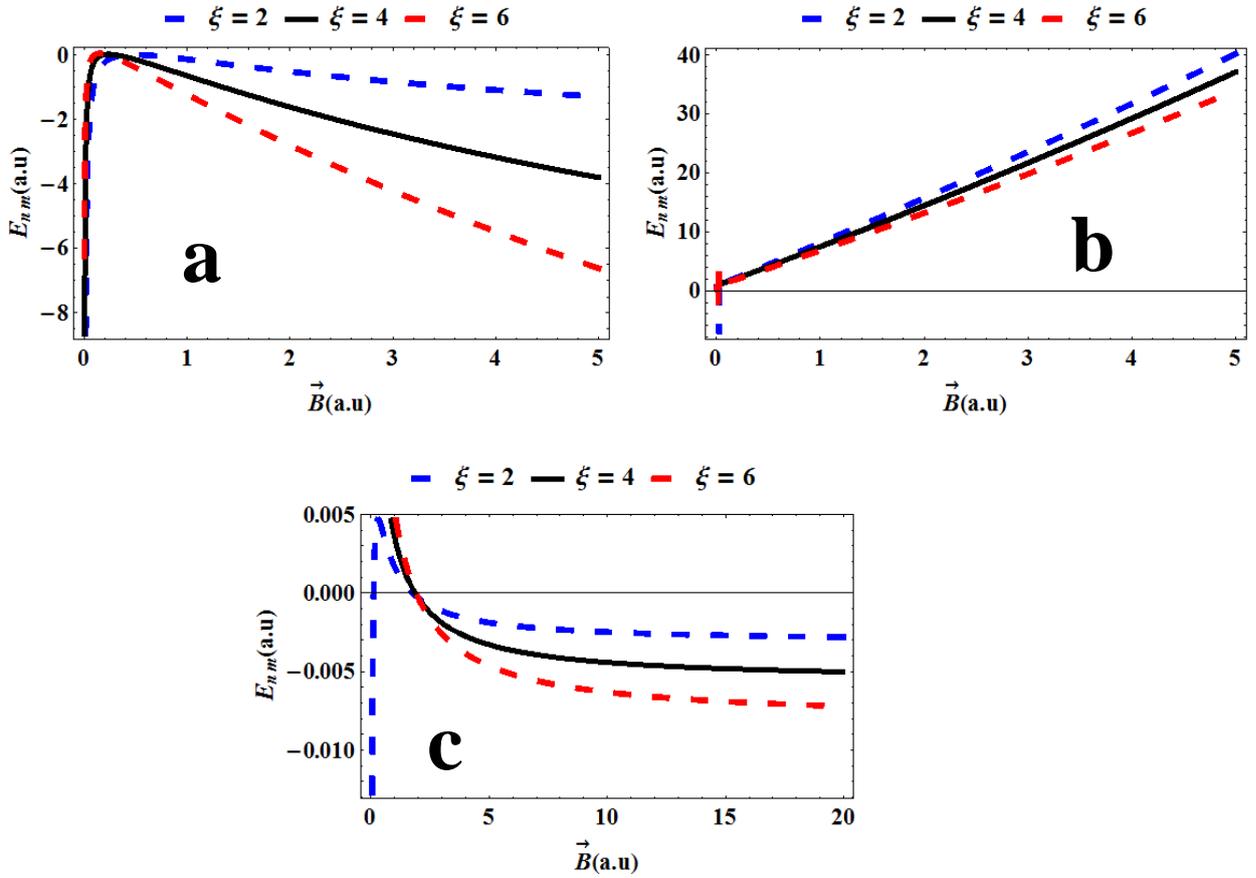

**FIG. 3.** Variation of energy values for the hydrogen atom in quantum plasmas and under the influence of the magnetic field and the AB flux field with topological defect in atomic units using the fitting parameters $m=n=0$ and $\lambda_D = 20$ **(a)** as a function of external magnetic field with various $\xi$ and $\eta = 0.04$. **(b)** Same as **(a)** but with $m=-1$ and $n=2$. **(c)** Same as **(a)** but with $\eta = 1$. All values are expressed in a.u.

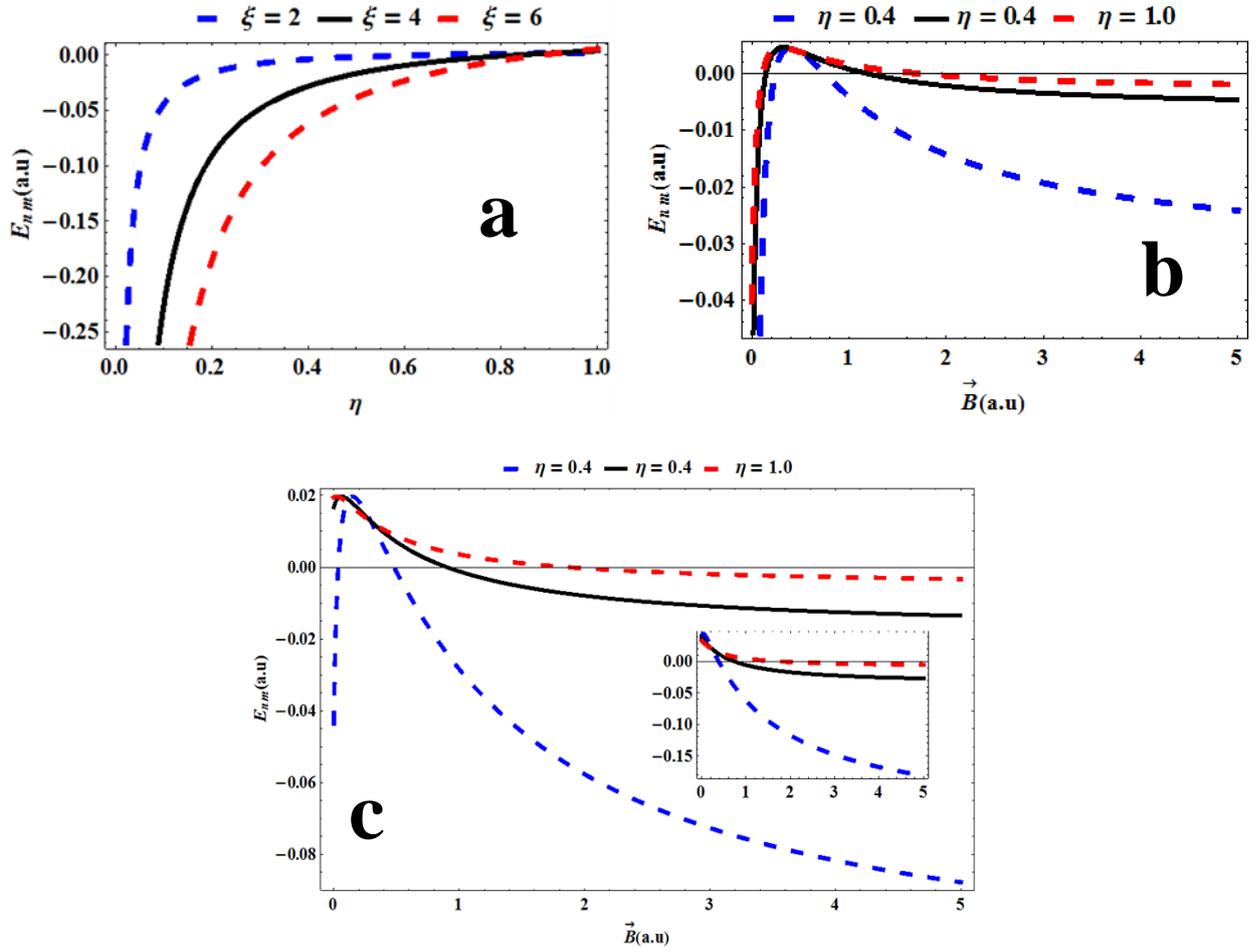

**FIG. 4.** Variation of energy values for the hydrogen atom in quantum plasmas and under the influence of the magnetic field and the AB flux field with topological defect in atomic units using the fitting parameters $m=n=0$ and $\lambda_D = 20$ **(a)** as a function of the topological defect with various $\xi$ and $\vec{B} = 1$. **(b)** as a function of external magnetic field with various $\eta$ and $\xi = 2$. **(c)** Same as **(b)** but with $\xi = 4$; the inset is for $\xi = 6$. All values are expressed in a.u.

.

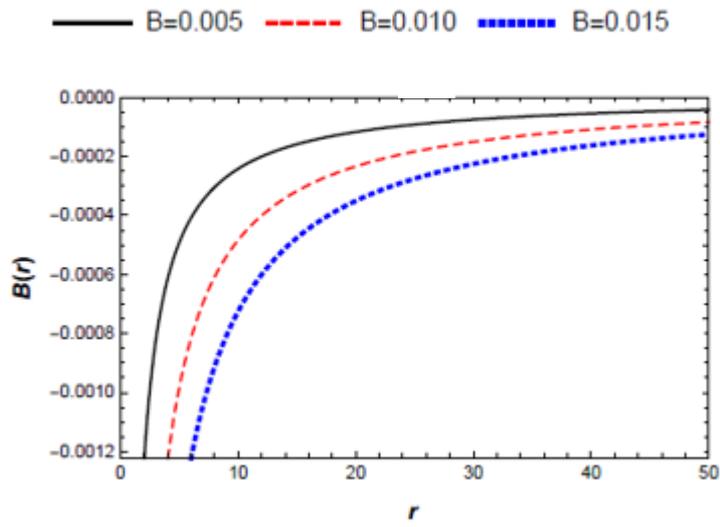

**Figure 5**: Magnetic field versus inter-particle distance.

**TABLE 1**. Energy values for the hydrogen atom in weakly coupled quantum plasma in the presence of topological defect and under the influence of AB flux and external magnetic fields with various values of magnetic quantum numbers. The following fitting parameters have been employed: $Z=1$, $\lambda_D = 20$, and note that $\eta = 1$ means absence of topological defect. All values are in a.u.

| m | n | $\vec{B} = \xi = 0, \eta = 1$ | $\vec{B} = 5, \xi = 0, \eta = 1$ | $\vec{B} = 0, \xi = 5, \eta = 1$ | $\vec{B} = \xi = 0, \eta = 0.005$ | $\vec{B} = \xi = 0, \eta = 0.5$ |
|---|---|---|---|---|---|---|
| 0 | 0 | -2.0003100 | -0.0004239 | 0.0286131 | -80000.0000000 | -8.0003100 |
|   | 1 | -0.2008680 | -0.0023947 | 0.0255455 | -8884.4500000 | -0.8453130 |
|   | 2 | -0.0581125 | -0.0067716 | 0.0215986 | -3195.2000000 | -0.2741130 |
|   | 3 | -0.0203125 | -0.0134862 | 0.0168847 | -1627.7600000 | -0.1182720 |
| 1 | 0 | -0.1892010 | -0.0011606 | 0.0340070 | 9.4772400 | -0.2403120 |
|   | 1 | -0.0535125 | -0.0055494 | 0.0296653 | 9.4304300 | -0.0998023 |
|   | 2 | -0.0176594 | -0.0122754 | 0.0246875 | 9.3818200 | -0.0437076 |
|   | 3 | -0.0046335 | -0.0212736 | 0.0190781 | 9.3314500 | -0.0171720 |
| -1 | 0 | -0.1892010 | 0.0003276 | 0.0194406 | 9.4772400 | -0.2403120 |
|    | 1 | -0.0535125 | 0.0008224 | 0.0193156 | 9.4304300 | -0.0998023 |
|    | 2 | -0.0176594 | -0.0011606 | 0.0172911 | 9.3818200 | -0.0437076 |
|    | 3 | -0.0046335 | -0.0055494 | 0.0139986 | 9.3314500 | -0.0171720 |

| m | n | $\vec{B} = 5, \xi = 0, \eta = 0.5$ | $\vec{B} = 0, \xi = 5, \eta = 0.5$ | $\vec{B} = \xi = 5, \eta = 1$ | $\vec{B} = \xi = 5, \eta = 0.5$ |
|---|---|---|---|---|---|
| 0 | 0 | -0.000325 | 0.024481 | -0.003968 | -0.080440 |
|   | 1 | -0.001807 | 0.029688 | -0.017579 | -0.102965 |
|   | 2 | -0.005710 | 0.030932 | -0.033278 | -0.127357 |
|   | 3 | -0.011965 | 0.029688 | -0.051010 | -0.153571 |
| 1 | 0 | -0.015141 | 0.057465 | -0.004637 | -0.146217 |
|   | 1 | -0.025820 | 0.053667 | -0.020448 | -0.175629 |
|   | 2 | -0.038665 | 0.048884 | -0.038289 | -0.206750 |
|   | 3 | -0.053618 | 0.043248 | -0.058110 | -0.239544 |
| -1 | 0 | -0.015550 | -0.077864 | -0.003286 | -0.0317934 |
|    | 1 | -0.006469 | -0.029201 | -0.014657 | -0.046653 |
|    | 2 | -0.000125 | -0.006428 | -0.028174 | -0.0635686 |
|    | 3 | 0.003565 | 0.004717 | -0.043780 | -0.0824881 |